\documentclass[12pt,preprint]{aastex}

\shorttitle{\emph{Swift}/XRT follow-up of TeV sources}
\shortauthors{Landi et al.}
\begin{document}

\title{\emph{Swift}/XRT follow-up observations of TeV sources of the HESS Inner Galaxy survey}

\author{R.~Landi\altaffilmark{1}, L.~Bassani\altaffilmark{1}, A.~Malizia\altaffilmark{1},
N.~Masetti\altaffilmark{1}, J.B.~Stephen\altaffilmark{1}, A.~Bazzano\altaffilmark{2},
P.~Ubertini\altaffilmark{2}, A.J.~Bird\altaffilmark{3}, and A.J.~Dean\altaffilmark{3}}

\altaffiltext{1}{INAF -- Istituto di Astrofisica Spaziale e Fisica Cosmica di Bologna, Via P. 
Gobetti 101, 40129 Bologna, Italy; landi@iasfbo.inaf.it}
\altaffiltext{2}{INAF -- Istituto di Astrofisica Spaziale e Fisica Cosmica di Roma, Via del 
Fosso del Cavaliere 100, 00133 Roma, Italy}
\altaffiltext{3}{School of Physics and Astronomy, University of Southampton, Highfield, 
     Southampton, SO17 1BJ, UK}

\begin{abstract}
In order to provide a firm identification of the newly discovered
Galactic TeV sources, a search for counterparts in a broad band from soft X-ray to soft gamma-rays 
is crucial as data in these wavebands allow us to distinguish between different types of suggested
models (for example leptonic versus hadronic) and, in turn, to disentangle their nature.
In this paper, we report the results of a set of follow-up observations performed by the
\emph{Swift}/X-Ray Telescope (XRT) on seven sources recently discovered by HESS, in the range from few 
hundred GeV to about 10 TeV, during the inner Galaxy survey (Aharonian et al. 2006). 
In all, but one case, we
detect X-ray sources inside or close-by the extended TeV emitting region. All these putative X-ray
counterparts have accurate arc-second location and are consistent with being point sources. The main
result of our search is the discovery that three of them are located at the center of the diffuse
radio emission of the supernova remnants, which have been spatially associated to these TeV objects.
HESS J1640--465, HESS J1834--087 and HESS J1813--178 show this evidence, suggestive of a possible
Pulsar Wind Nebula association.
\end{abstract}


\section{Introduction}

The HESS (High Energy Stereoscopic System) collaboration has recently
reported results of a first sensitive survey of the inner part of our galaxy in very high energy 
gamma-rays (Aharonian et al. 2005, 2006). This survey has revealed the existence of a 
population of 
fourteen TeV objects, most of which were previously unknown. These
findings have important astrophysical implications for our understanding of 
cosmic particle accelerators via TeV measurements. 
Various types of sources in
the galaxy can act as cosmic accelerators: pulsars and their pulsar
wind nebulae (PWN), supernova remnants (SNR), star forming regions and possibly
binary systems containing a collapsed object such as a microquasar or a pulsar.
Indeed, four of these new TeV sources (HESS J1640--465, HESS J1713--381, HESS J1813--178 and 
HESS J1834--087) are spatially coincident with SNRs, which have been mapped and studied 
at radio frequencies (Green et al. 1999; Green 2004; Brogan et al. 2005; Helfand et al. 
2005); in the first three cases X-ray radiation associated with the 
radio emission has also been reported (Sugizaki et al. 2001; Aharonian et al. 2006; Ubertini et al. 
2005), while in the last case no X-ray data has so far been published.
Three more sources, HESS J1616--509, HESS J1804--216 and HESS J1825--137, 
could be associated with energetic nearby pulsars, but their spatial offset from the TeV
emitting region is too large to make the association secure.
Other three objects are consistent with the 95\% positional error contour of 
unidentified EGRET sources (HESS J1640--465, HESS J1745--303 and HESS J1825--137)
(Hartman et al. 1999), while  
two are located very close to a new class of highly absorbed 
hard X-ray binaries recently detected by \emph{INTEGRAL} (HESS J1632--478 and HESS J1634--472, Walter et 
al. 2006; Lutovinov et al. 2006). 
Another source (HESS J1837--069) is related to X-ray emission of unknown origin 
detected by \emph{ASCA} and \emph{Beppo}SAX/\emph{INTEGRAL} (Malizia et al. 2005) and may either be 
associated to a SNR/PWN or belong to the class of absorbed hard X-ray binaries. 
The three remaining objects (HESS J1614--518, HESS J1702--420 and HESS J1708--410) 
have no plausible SNR, pulsar, or EGRET counterpart and therefore it has been suggested in the literature 
that they belong to a new source type (Aharonian et al. 2005).
However, it should also be remembered that the probability of chance coincidence in the crowded
region of the galactic plane is quite high and therefore also for the other sources the reported 
associations must be verified carefully.

In the case of HESS J1813--178 and HESS J1837--069, the existing X/soft gamma-ray data
support at first glance the
Synchrotron/Inverse Compton scenario (Malizia et al. 2005; Ubertini et al. 2005), but, unfortunately, 
various attempts to model in detail their overall spectral energy distribution (Brogan et al. 
2006; Albert et al. 2006) have not been completely successful (Ubertini 2006).

The lack of radio/X-ray emission from some of the HESS sources is particularly
interesting as it strongly suggests that the accelerated particles
may be  nucleons rather than high energy electrons. However, the possibility that still
undetected SNRs or PWNs are present inside  the TeV emitting region must be considered since 
a deep multiwavelength investigation 
of each HESS source is still lacking. Furthermore, the localization of any suggested 
X-ray/radio counterpart with sufficient precision to 
allow follow-up observations (for example at optical/infrared frequencies) is crucial in order to obtain
secure identification and, in turn, to understand the nature of the TeV 
emission. In this paper, we report on follow-up X-ray observations of seven of these HESS survey 
sources 
with the \emph{Swift}/XRT instrument. This study provides a list of X-ray objects detected inside 
(or in the vicinity of) the TeV emitting region, their accurate arc-second location and information on 
their counterparts at other wavebands. For the brightest 
sources, spectral information is also obtained in the X-ray band.

\section{Data analysis and results}

To search for X-ray counterparts of HESS survey sources we have used data collected 
with the XRT (X-ray Telescope, 0.2--10 keV) on board the \emph{Swift} satellite (Gehrels et al.
2004). The log of all observations analyzed in the present paper is given in 
Table 1, where we report for each measurement the observation date, the pointing coordinates and
the exposure time. 
Data reduction was performed using the XRTDAS v2.4 standard data pipeline package
({\sc xrtpipeline} v. 0.10.3), in order to
produce screened event files. All data were extracted only in the Photon Counting
(PC) mode (Hill et al. 2004),
adopting the standard grade filtering (0--12 for PC) according to the XRT nomenclature.
For each observation an image in the 0.3--10 keV band was obtained and analyzed. Sources detected 
above 3$\sigma$ confidence level are listed in Table 2, where we report for each detection its location, 
count rate (0.2--10 keV), error box radius and 2--10 keV flux corrected for absorption.

Only in one case, that of HESS J1303--631, we do not detect any source in the XRT image; this is in 
substantial 
agreement with \emph{Chandra} results from Mukherjee \& Halpern (2005) since the only source they 
reported is too weak (2--10 keV flux of $\sim$4 $\times10^{-14}$ erg cm$^{-2}$ s$^{-1}$) for a meaningful 
detection by XRT in a short exposure (4.7 ks).
In the case of HESS J1813--178, 
the image was corrupted (see dedicated section), but this did not prevent 
the localization and analysis of the only X-ray source present in the field.
  
Each X-ray source detected has been tested against the instrument point spread function, in order
to assess if it was point-like or diffuse: none of the detected sources has a 
profile indicative of diffuse emission. Only for the most interesting objects has a refined position and 
error 
box been estimated using the method reported in Moretti et al. (2006).
The radius of the error boxes associated with these XRT detections ranges from
3$^{\prime \prime}$ to 6$^{\prime \prime}$ (see Table 2).
Within these error boxes, we have 
searched the HEASARC data base for optical/infrared counterparts of the most interesting objects.
Given the spatial association of four HESS sources (HESS J1640--465, HESS J1804--216, HESS J1813--178 
and HESS J1834--087) with SNRs and the possibility of using radio images 
from the NVSS (National Radio Astronomy Observatory (NRAO) Very Large Array (VLA) Sky Survey Catalog, 
Condon et al. 1998) or
MOST (Molonglo Observatory Synthesis Telescope, Robertson 1991) to map their emission, 
we indicate the location of all X-ray sources detected in these four cases on the radio 
maps (Figure 2, 3, 4 and 5); in these radio images, the green circles indicate the position and size
of the associated SNR as reported in Green (2004), Brogan et al. (2005) and Brogan et al. (2006).
In the remaining two cases (HESS J1614--518 and HESS J1837--069), XRT images are
presented instead (Figure 1 and 6). In all figures, the red circle/ellipse represents the extension of the 
TeV source 
as reported by Aharonian et al. (2006); X-ray sources are instead indicated by a box and/or a number
as listed in Table 2. 

Source spectral data were
extracted using photons from a circular region of radius 20$^{\prime \prime}$; background spectra come
from various uncontaminated regions near the X-ray source using  
a circular/annular region of various radius size depending on the presence of contaminating sources
within the XRT field of view.
In all cases, the spectra were extracted from the events file using {\sc XSELECT} software and
binned using {\sc grppha} in an appropriate way, so 
that the $\chi^{2}$ statistic could be used reliably.
We used the lastest version (v.008) of the response matrices and create individual ancillary
response files (ARF) using {\sc xrtmkarf}.
Spectral analyses have been performed using XSPEC version 12.2.1.
Due to the limited statistical quality of most spectra, we simply employed a power law absorbed by 
both a Galactic (Dickey \& Lockman 1990) and, when applicable, an intrinsic column density.
In most cases only 2--10 keV fluxes have been estimated assuming a Crab-like spectrum (see Table 2).
For a few sources with sufficient statistics ($\geq$ 5$\sigma$), we have performed a proper spectral 
analysis and the results are presented in Table 3.
All quoted errors correspond to a 90$\%$ confidence level 
for one interesting parameter ($\Delta\chi^{2}=2.71$).
Unfortunately, the observational mode of XRT does not allow continuous monitoring of the source;
this, coupled with the low statistical quality of the data, makes a timing analysis unpractical.

\subsection{HESS J1614--518}

HESS J1614--518 is characterized by an extended and elongated morphology.
This source, which is one of the two brightest TeV objects detected in the HESS survey, 
is located in a region 
relatively poor in counterpart candidates. A \emph{Chandra} observation of the nearby sky 
(10$^{\prime}$ away) 
did not show evidence for X-ray emission (Kastner et al. 2003). 
The XRT observation, which instead covers the entire 
HESS emitting region (see Figure 1), indicates that only two faint X-ray sources are present 
inside or close to the HESS ellipse.
Due to the weakness of both sources, no spectral analysis is possible and only a 
flux has been extracted.
Source 1 has a USNO-B1.0 (Monet et al. 1999) and 2MASS (Two Micron All Sky Survey, 
Skrutskie et al. 2006) counterpart at R.A.=16$^{h}$14$^{m}$ 06$^{s}$.10 and 
Dec=--51$^{\circ}$ 52$^{\prime}$ 26$^{\prime \prime}$.4 (J2000), with $K$ $\simeq$9.9 and $R$ 
$\simeq$12.5;
the $R-K$ colour index of $\sim$3 implies extinction in the source direction.
Source 2 is likely to be associated with a RASS (\emph{ROSAT} All Sky Survey) faint source 
(RXS J161319.8--514329) and has nearby (42$^{\prime \prime}$) 
a very bright and probably extended radio source (PMN J1613--5143) characterized by a flux of 
3.2 Jy at 4850 MHz (Wright et al. 1994); also in this case a USNO-B1.0/2MASS source is a 
possible counterpart of the XRT object with coordinates R.A.=16$^{h}$ 13$^{m}$ 20$^{s}$.85 and 
Dec=--51$^{\circ}$ 43$^{\prime}$ 17$^{\prime \prime}$.4 (J2000) and optical/infrared 
magnitudes of $R$ $\simeq$11.6 and $K$ $\simeq$8.4.
The presence of such a bright radio source near object 2 and at the border of the TeV emitting 
region is intriguing and we encourage further 
observations of PMN J1613--5143, in order to understand its nature and possible association with 
HESS J1614--518.

\subsection{HESS J1640--465}

HESS J1640--465 is marginally extended at TeV energies and quasi-symmetrical in shape.
The most promising counterpart is the broken shell SNR (G338.3--0.0) also detected by 
\emph{ASCA} (AX J1640--4632) with a 0.7--10 keV flux of $\sim$$1.2\times10^{-12}$ erg cm$^{-2}$ s$^{-1}$
and a soft ($\Gamma=2.98^{+1.13}_{-0.89}$) and absorbed 
($N_{\rm H}=9.6^{+4.7}_{-3.3} \times10^{22}$ cm$^{-2}$) X-ray spectrum typical of SNR hot plasma 
emission (Sugizaki et al. 2001).
The MOST 843 MHz radio image of this region (see Figure 2) provides insight into this SNR clearly showing 
a broken shell-like structure. 
G338.3--0.0 is 8$^{\prime}$ in diameter and is located in a complex region at 
8.6 kpc distance (Guseinov, Ankay $\&$ Tagieva 2004);
the larger circle in Figure 2, which should provide its location and size, is slightly offset
with respect to the SNR, but this may be due to uncertainties in the source parameters and also to
differences in images taken at different radio frequencies. Despite this, it is unquestionable 
that the HESS extension matches the SNR structure quite closely. G338.3--0.0 is not very well 
studied as the
radio spectral index is unknown and the 1 GHz flux is set around 7 Jy with great uncertainty
(Green 2004).
Only one (Source 1) of the five objects detected by XRT is inside the TeV extension.
Analysis of the X-ray spectrum of this source shows that it is 
fully compatible with that found by \emph{ASCA} (see Table 3), although an 
equally 
good fit is obtained with a flat spectrum ($\Gamma =0.65^{+0.55}_{-0.84}$), no absorption and 
a similar flux. This source is clearly the object detected by \emph{ASCA}, but it is seen as point-like 
by XRT and localized with arc-second precision.
Source 3, which is the brightest X-ray source in the region, is very soft, being undetected 
above 3 keV and characterized by a thermal bremsstrahlung emission with $kT$ $\sim$0.43, while  
XRT data of Source 2 are well described by an absorbed power law ($N_{\rm H}$ 
$\sim$$3.4\times10^{22}$
cm$^{-2}$) with photon index $\Gamma$ $\sim$1.8 (see Table 3 for both objects).

For the remaining two sources, the data are of too low statistical quality to allow 
spectral analysis and only a flux has been obtained. 
It is clear that the most interesting object coming out of this XRT observation is Source 1;
unfortunately, within its positional uncertainty we do not find any cataloged object in the HEASARC 
archive, implying a very weak source at optical/infrared wavelengths.

\subsection{HESS J1804--216}

HESS J1804--216 is an extended source and it is spatially coincident with the south western part
of SNR G8.7--0.1, which is located in the W30 complex. The morphology of this remnant is 
neither shell-like nor center-filled with "plumes" observed in the south part and along the western 
edge; it has a radius of 22$^{\prime}$.5, a radio spectral index $\alpha=0.5$ ($S_{\nu}$ 
$\sim$ $\nu^{-\alpha}$)
and a 1 GHz flux of 80 Jy (Green 2004). This SNR may also be linked with the young 
radio pulsar PSR J1803--2137.
Both objects were observed by \emph{ROSAT}/IPC and detected in soft X-rays: the pulsar 
is weak and characterized by a hard 
and absorbed spectrum, while the X-ray emission of G8.7--0.1 is brighter and diffuse (Finley \&
{\"O}gelman 1994). 
Very recently, Brogan et al. (2006) reported the discovery of a
new SNR, which is located just inside the TeV extension (G8.31--0.9); 
it is a small ($5^{\prime}$$\times$$4^{\prime}$), shell-type SNR with a radio spectral index $\alpha$ in 
the range 0.6--0.7 and a 1 GHz flux around 1--2 Jy. 
The NVSS map of this region is shown in Figure 3,
where we outline the two SNRs, as well as the three X-ray sources detected by XRT: only two
are located well inside the TeV extension. 
Sources 1 and 2 have
already been detected in X-rays by \emph{ROSAT} as a RASS faint source (\# 1, 1RXS J180404.6-215325) 
and WGCAT/IPC object (\# 2). Source 3 has not been seen previously, but it is the one closest
to a radio complex of unknown nature (around 30 mJy in flux in the NVSS catalog),
which is possibly part of the W30 complex and it is just barely visible in the figure.
Within the positional uncertainty of this XRT source, we do not find any optical and/or infrared 
counterpart in the various HEASARC catalogs. 
The XRT spectrum of Source 1 is very soft, well described by thermal emission from optically thin 
plasma (Raymond \& Smith 1977), and has no absorption in excess to the galactic one (see Table 3), 
suggesting that it might originate from a bright star. Indeed, within the XRT positional uncertainty, 
an object is found in the USNO-B1.0/2MASS catalogs with coordinates R.A.=18$^{h}$ 04$^{m}$ 03$^{s}$.23 and
Dec=--21$^{\circ}$ 53$^{\prime}$ 36$^{\prime \prime}$.5 (J2000) and optical/infrared
magnitudes of $R$ $\simeq$12.1 and $K$ $\simeq$9.3.
The other two objects (Source 2 and 3) are not bright enough in X-rays to allow spectral 
analysis and only fluxes were extracted from the XRT data.
Within their positional uncertainties, we do not find any counterparts; however, if we 
enlarge the error box of Source 2 to $6^{\prime \prime}$, then a counterpart in a 
USNO-B1.0/2MASS catalogs is found at R.A.=18$^{h}$ 04$^{m}$ 00$^{s}$.72 and
Dec=--21$^{\circ}$ 42$^{\prime}$ 52$^{\prime \prime}$.4 (J2000), with magnitude $R$ $\simeq$13.2 
and $K$ $\simeq$10.2.

\subsection{HESS J1813--178}

HESS J1813--178 has been the target of various studies due to the initial lack of plausible
counterparts (Aharonian et al. 2005). Very recently, the MAGIC 
experiment has confirmed the HESS detection (Albert et al. 2006).
This source is one of the less extended TeV objects and has a 
spatial coincidence with a shell type supernova (G12.82--0.02) recently detected in radio
(Brogan et al. 2005, Helfand, Becker \& White 2005).
This SNR has a small extension (2$^{\prime}$.5 in diameter), a radio spectral 
index $\alpha=0.5$ and a 1 GHz flux of 0.16 Jy.
The HESS source is also associated with an X-ray/gamma-ray emitting object (AX J1813--178, IGR 
J18135--1751) fairly bright in both  
these wavebands (Ubertini et al. 2005). However, the X-ray/gamma-ray error boxes are too large 
($1^{\prime}$--$3^{\prime}$) to allow a proper localization of this source and therefore any 
relationship with the SNR.
The XRT image is highly contaminated by the presence of fringes in the image: these are likely due 
to the presence of a very bright source (GX 13+1) outside the XRT image 
(the same problem that affected the \emph{ASCA} observation; see Ubertini et al. 2005).
In any case, a point-like source, compatible with the \emph{ASCA}/\emph{INTEGRAL} error boxes, is 
clearly 
visible in both observations available for this region. The XRT position is consistent with being the 
same in both measurements, suggesting that it is a real feature
rather than an artifact due to the fringes. 
The HESS extension is located 
within the SNR emitting region, and the XRT source lies within it, i.e. it is almost at the center
of the shell (see Figure 4, which is a map from the NVSS at 20 cm).
 
The XRT spectrum of this source is comparable in spectral parameters with the \emph{ASCA} one (see 
Table 3), but the flux is a factor of 2 lower. Instead, 
we do not find any evidence for flux variation between the two XRT observations. 
The XRT point source has a possible counterpart in the 2MASS/DENIS (Deep Near Infrared Survey
of the Southern Sky, Paturel et al. 2003) catalogs at
R.A.=18$^{h}$ 13$^{m}$ 35$^{s}$.06 and Dec=--17$^{\circ}$ 49$^{\prime}$ 
52$^{\prime \prime}$.4 (J2000); this object has $\simeq$11.8 and $\simeq$16.7 magnitudes in the K and 
I bands respectively
and it is not listed in the USNO-B1.0 catalog, again indicative of strong extinction in the source 
direction.

\subsection{HESS J1834--087}

HESS J1834--087 is an extended TeV source which has a positional coincidence 
with the SNR G23.3--0.3 (W41) characterized by a size of 27$^{\prime}$ in diameter, and
located at an estimated distance of 4.8 kpc. Its radio spectral index $\alpha$ is 0.5 similar to the
other SNRs, while the 1 GHz flux is 70 Jy (Green 2004). The 20 cm NVSS radio image (Figure 5)
clearly shows the SNR broken shell with a hot spot at its center. The TeV source is fully
located within the shell structure and it is well associated with the central hot spot.
Very recently, the MAGIC collaboration reported the detection of HESS J1834--087 and confirmed the
extended nature of the TeV emission (Albert et al. 2006).
No X-ray data are yet available from this region.
The present XRT observation pinpoints for the first time a faint X-ray source (Source 1) detected
inside the HESS extension and located close to the hot spot at the center of the SNR.
Another two X-ray objects of similar brightness lie just outside the HESS extension and 
only one of the two (Source 2) corresponds to a hot spot in the SNR shell. 
All sources are too weak for spectral analysis and only an indication of the X-ray flux can be derived.
Within the XRT positional uncertainty of Source 1, we find an optical counterpart belonging 
to the USNO-B1.0 catalog and located at     
R.A.=18$^{h}$ 34$^{m}$ 34$^{s}$.90 and  
Dec=--08$^{\circ}$ 44$^{\prime}$ 49$^{\prime \prime}$.6 (J2000), with magnitude $R$ $\simeq$18.3.
This source, also listed in the 2MASS catalog with $K$ $\simeq$13 magnitude, is thus heavily reddened
having an $R-K$ color index of $\sim$5.

\subsection{HESS J1837--069}

This HESS source has an elongated shape and overlaps the southern part of 
the region of diffuse X-ray emission seen by \emph{ASCA}
(Bamba et al. 2003, Figure 1b);
XRT does not detect any diffuse emission, but locates the same (within the instrumental 
accuracy) point sources identified by \emph{ASCA}\footnote{Source 8 in Bamba et al. (2003) appears 
to have the wrong coordinates in their Table 5, but its location in their Figure 1b agrees with our 
position.}.
The northern part of the \emph{ASCA} region, outside the XRT field of view, corresponds to G25.5+0.0,
whose nature is still unknown, although the X-ray emission indicates
that it may be a shell-type supernova or a pulsar wind nebula. 
The brightest feature in our image (Source 1) is the only source detected above 10 keV both by
the \emph{INTEGRAL}/IBIS
and \emph{Beppo}SAX/PDS instruments (Malizia et al. 2005), but it is located just outside the HESS 
extension (see, however, Aharonian et al. 2005 for a discussion on its possible association with 
the TeV object).
Despite the good localization provided by the \emph{Einstein}/IPC
instrument, the error box associated with this X/gamma-ray source has up to now been too large to
provide a unique optical counterpart. 
Source 1 is fully compatible with the \emph{ASCA}/\emph{INTEGRAL} error boxes, but the XRT localization 
accuracy allows its positional uncertainty to be considerably improved and its
counterparts to be searched at other wavebands. The XRT spectrum is slightly steeper than the 
\emph{ASCA} one (see Table 3), but compatible with the combined \emph{ASCA}/\emph{INTEGRAL} 
observed shape (Malizia et al. 2005). Only one source is found within the
XRT error box from the USNO-B1.0/2MASS catalogs (see Table 4); it has
$R-K$ $\sim$6.7, again suggesting extreme reddening.
For the remaining objects detected within the XRT field of view, we find that five have an 
optical/infrared counterpart (see Table 4); of these objects only one (Source 7) is located inside the HESS 
ellipse.
Unfortunately, sources from 2 to 9 are too faint to allow a proper spectral analysis of the 
XRT data and only fluxes are reported in Table 2.
 
\section{Discussion and Conclusions}

To provide a firm identification of lower energy counterparts to the large number of newly discovered HESS 
sources, we have searched in the TeV emitting region for the presence of \emph{Swift}/XRT counterparts.
For six of them (see Section 2.1 to 2.6) the search has been successful, with one or more X-ray sources 
accurately positioned within or in the close vicinity of the HESS extension.
However, before claiming that the same object or emission process is responsible for the X and 
gamma-ray emission, an assessment of the correctness
of these associations must be performed.
In particular, the random chance probability of finding an X-ray source within the HESS extension
must be properly evaluated.
We can do this using the galactic LogN-LogS relation obtained by \emph{ASCA} in the 2--10 keV energy 
band (Yamauchi et al. 2002), always with the caveat that this assumes a uniform distribution
of independent sources.
Out of the whole TeV sample, only in the case of 
HESS J1804--216, do we find that the two XRT objects detected are very likely due to 
random coincidence, while in all other cases the probability of finding an X-ray source
inside the TeV extension by chance is around 0.01
for HESS J1640--465 and HESS J1813--178, 0.2 for HESS J1614--518 and 0.4 for HESS J1834--087.
The chance probability that the X-ray sources are background AGNs is even lower.
We therefore consider the X-ray sources discovered in the first two cases 
as very likely counterparts of the TeV objects.
In the remaining two cases, the X-ray sources are just interesting objects, which need a more
in depth analysis.
For the case of HESS J1837--069, even though AX J1838.0--0655 is located slightly outside 
the (1$\sigma$) HESS extension, it is considered the most promising candidate counterpart 
(Aharonian et al. 2005, 2006), although any firm conclusion on their association must await
optical follow-up spectroscopy; this is now possible with XRT location. 

An important result of the present study is also the arc-second localization
of all X-ray sources, in particular for the most likely candidates.
In fact, the precise localization allows in most cases to pinpoint the optical/infrared counterpart of the 
X-ray source, thus giving the possibility to assess its nature by means of follow-up observations,
which will hopefully help establishing their association with the TeV object.
The main outcome of our follow-up work at X-ray wavelengths is the discovery that three of the detected 
X-ray sources are located 
at the center of a SNR, which is spatially coincidence with the TeV object: at least for
HESS J1640--465, HESS J1834--087 and
HESS J1813--178 this evidence is clear in the radio maps (see Figure 2, 4 and 5) and suggestive of a pulsar 
wind Nebula type of association.

So far, because of the limited time span and statistics of the present X-ray observations, it has been 
impossible to detect the timing signature of a pulsar in the detected objects.
Future work will therefore need to concentrate on more in depth studies of these sources at X-ray 
energies and on optical/infrared follow-up observations of their counterparts if we want to understand 
their nature and connection to the TeV emission.

\section{Acknowledgements}
This research has been supported by ASI under contracts I/R/046/04 and I/023/05. 
This research has made use of data obtained from SIMBAD
(CDS, Strasbourg, France) and HEASARC (NASA's Goddard Space Flight Center).
We whish to thank the anonymous referee for his/her comments and suggestions, which have
improved the quality of this work.

\clearpage

\begin{deluxetable}{lcccc}
\tablecolumns{5}
\tablecaption{\emph{Swift}/XRT Observation Log.}
\startdata
\hline
\hline
Name  &\multicolumn{2}{c}{Pointing Coordinates} &  Observation date & Exposure \\
    &   R.A.(J2000) &  Dec(J2000) &     &    (sec) \\
\hline
\hline
HESS J1303--631 & 13$^{h}$ 02$^{m}$ 23$^{s}$.9 & --63$^{\circ}$ 11$^{\prime}$ 42$^{\prime \prime}$.4
& Nov 15, 2005   &  4665 \\
\hline
HESS J1614--518  & 16$^{h}$ 14$^{m}$ 15$^{s}$.0 & --51$^{\circ}$ 49$^{\prime}$ 08$^{\prime \prime}$.7
& Mar 17, 2006   &  1666 \\
\hline
HESS J1640--465  & 16$^{h}$ 40$^{m}$ 37$^{s}$.5 & --46$^{\circ}$ 31$^{\prime}$ 36$^{\prime \prime}$.2
& Feb 4, 2006   & 2430 \\
HESS J1640--465  & 16$^{h}$ 40$^{m}$ 28$^{s}$.6 & --46$^{\circ}$ 32$^{\prime}$ 33$^{\prime \prime}$.4
& Mar 15, 2006   & 15008 \\
\hline
HESS J1804--216  & 18$^{h}$ 04$^{m}$ 29$^{s}$.7 & --21$^{\circ}$ 43$^{\prime}$ 57$^{\prime \prime}$.2
& Nov 3, 2005    & 11624 \\
\hline
HESS J1813--178  & 18$^{h}$ 13$^{m}$ 37$^{s}$.9 & --17$^{\circ}$ 52$^{\prime}$ 03$^{\prime \prime}$.7
& Nov 3, 2005    &  3229 \\
HESS J1813--178  & 18$^{h}$ 13$^{m}$ 30$^{s}$.1 & --17$^{\circ}$ 51$^{\prime}$ 19$^{\prime \prime}$.4
& Mar 9, 2006    &  7204 \\
\hline
HESS J1834--087  & 18$^{h}$ 34$^{m}$ 51$^{s}$.8 & --08$^{\circ}$ 44$^{\prime}$ 38$^{\prime \prime}$.2
& Jun 29, 2005   &  6917 \\
\hline
HESS J1837--069  & 18$^{h}$ 37$^{m}$ 31$^{s}$.6 & --06$^{\circ}$ 57$^{\prime}$ 17$^{\prime \prime}$.7
& Mar 3, 2006    & 11868 \\
\hline
\enddata
\end{deluxetable}

\clearpage

\begin{deluxetable}{lccccc}
\tablecolumns{6}
\tabletypesize{\scriptsize}
\tablecaption{Sources detected in the XRT field of view of the HESS pointing}
\startdata
\hline
\hline
 Source & R.A.(J2000) & Dec(J2000) & Count rate$^{a,\dag}$ &  Error box & Flux$^{b,c,\dag}$ \\
  &   &   &   (0.2--10 keV)& (arcsec) & (2--10 keV) \\
\hline
\hline
\multicolumn{6}{c}{HESS J1614--518} \\
\hline
\# 1 & 16$^{h}$14$^{m}$05$^{s}$.8 & --51$^{\circ}$52$^{\prime}$26$^{\prime \prime}$.0 &
6.61$\pm$2.11 &  6.0  &  0.54$^{c}$ \\
\# 2 & 16$^{h}$13$^{m}$20$^{s}$.4 & --51$^{\circ}$43$^{\prime}$17$^{\prime \prime}$.0 &
5.71$\pm$1.75 & 6.0  & 0.66$^{c}$ \\
\hline
\multicolumn{6}{c}{HESS J1640--465} \\
\hline
\# 1 & 16$^{h}$40$^{m}$43$^{s}$.5 & --46$^{\circ}$31$^{\prime}$38$^{\prime \prime}$.6 &
4.64$\pm$0.68 & 4.1  &  0.72 \\
\# 2 & 16$^{h}$40$^{m}$29$^{s}$.2 & --46$^{\circ}$23$^{\prime}$29$^{\prime \prime}$.0 &
3.51$\pm$0.60 & 6.0  & 0.68  \\
\# 3 & 16$^{h}$41$^{m}$30$^{s}$.5 & --46$^{\circ}$30$^{\prime}$47$^{\prime \prime}$.8 &
6.08$\pm$0.73 & 6.0  & 0.014  \\
\# 4 & 16$^{h}$41$^{m}$14$^{s}$.5 & --46$^{\circ}$31$^{\prime}$29$^{\prime \prime}$.2 &
0.98$\pm$0.30 & 6.0  & 0.057$^{c}$  \\
\# 5 & 16$^{h}$41$^{m}$16$^{s}$.0 & --46$^{\circ}$32$^{\prime}$29$^{\prime \prime}$.0 &
0.73$\pm$0.24 & 6.0  & 0.086$^{c}$  \\
\hline
\multicolumn{6}{c}{HESS J1804--216} \\
\hline
\# 1 & 18$^{h}$04$^{m}$03$^{s}$.2 & --21$^{\circ}$53$^{\prime}$31$^{\prime \prime}$.5 &
12.90$\pm$1.12 & 6.0  & 0.04  \\
\# 2 & 18$^{h}$04$^{m}$00$^{s}$.8 & --21$^{\circ}$42$^{\prime}$47$^{\prime \prime}$.5 &
1.87$\pm$0.57 & 4.9  & 0.14$^{c}$  \\
\# 3 & 18$^{h}$04$^{m}$32$^{s}$.6 & --21$^{\circ}$40$^{\prime}$01$^{\prime \prime}$.8 &
2.24$\pm$0.54 & 5.0  & 0.22$^{c}$  \\
\hline
\multicolumn{6}{c}{HESS J1813--178} \\
\hline
\# 1 & 18$^{h}$13$^{m}$34$^{s}$.9 & --17$^{\circ}$49$^{\prime}$53$^{\prime \prime}$.2 &
35.4$\pm$2.5 & 3.5  &  0.98 \\
\hline
\multicolumn{5}{c}{HESS J1834--087} \\
\hline
\# 1 & 18$^{h}$34$^{m}$34$^{s}$.7 & --08$^{\circ}$44$^{\prime}$45$^{\prime \prime}$.9 &
2.13$\pm$0.70 & 6.0  & 0.26$^{c}$  \\
\# 2 & 18$^{h}$34$^{m}$07$^{s}$.2 & --08$^{\circ}$51$^{\prime}$59$^{\prime \prime}$.0 &
2.34$\pm$0.70 & 6.0 &  9.9$^{c}$    \\
\# 3 & 18$^{h}$35$^{m}$14$^{s}$.5 & --08$^{\circ}$37$^{\prime}$46$^{\prime \prime}$.2 &
1.95$\pm$0.63 & 6.0   &  0.18$^{c}$ \\
\hline
\multicolumn{6}{c}{HESS J1837--069} \\
\hline
\# 1 & 18$^{h}$38$^{m}$03$^{s}$.1 & --06$^{\circ}$55$^{\prime}$38$^{\prime \prime}$.6 &
42.8$\pm$2.0 & 3.4  & 11.3  \\
\# 2 & 18$^{h}$37$^{m}$49$^{s}$.7 & --07$^{\circ}$07$^{\prime}$33$^{\prime \prime}$.2 &
1.25$\pm$0.41 & 6.0   & 0.13$^{c}$ \\
\# 3 & 18$^{h}$37$^{m}$41$^{s}$.4 & --07$^{\circ}$05$^{\prime}$16$^{\prime \prime}$.4 &
1.36$\pm$0.41 & 6.0  & 0.14$^{c}$  \\
\# 4 & 18$^{h}$36$^{m}$48$^{s}$.5 & --07$^{\circ}$04$^{\prime}$44$^{\prime \prime}$.0 &
2.15$\pm$0.55 & 6.0  & 0.13$^{c}$ \\
\# 5 & 18$^{h}$37$^{m}$02$^{s}$.5 & --06$^{\circ}$49$^{\prime}$28$^{\prime \prime}$.5 &
1.80$\pm$0.50 & 6.0   & 0.19$^{c}$ \\
\# 6 & 18$^{h}$37$^{m}$20$^{s}$.7 & --06$^{\circ}$52$^{\prime}$47$^{\prime \prime}$.2 &
2.18$\pm$0.44 & 6.0  & 0.38$^{c}$ \\
\# 7 & 18$^{h}$37$^{m}$23$^{s}$.0 & --06$^{\circ}$52$^{\prime}$13$^{\prime \prime}$.4 &
1.28$\pm$0.36 & 6.0  &  0.18$^{c}$ \\
\# 8 & 18$^{h}$37$^{m}$59$^{s}$.4 & --06$^{\circ}$49$^{\prime}$23$^{\prime \prime}$.8 &
2.68$\pm$0.58 & 6.0  &  0.36$^{c}$ \\
\# 9 & 18$^{h}$37$^{m}$51$^{s}$.3 & --06$^{\circ}$53$^{\prime}$48$^{\prime \prime}$.4 &
2.32$\pm$0.55 & 6.0 & 0.21$^{c}$  \\
\hline
\enddata

\tablecomments{
$^{a}$ In units of $10^{-3}$ counts s$^{-1}$;\\
$^{b}$ In units of $10^{-12}$ erg cm$^{-2}$ s$^{-1}$;\\
$^{c}$ For these sources the 2--10 keV flux has been computed assuming a Crab-like spectrum;\\
$^{\dagger}$ The relation between 0.2--10 keV count rate and 2--10 keV flux may differ from source to 
source because of two effects: source off-axis position (vignetting) and source spectral shape.
}

\end{deluxetable}

\clearpage

\begin{deluxetable}{lcccccc}
\tablecolumns{7}
\tabletypesize{\scriptsize}
\tablecaption{XRT spectral analysis results.}
\startdata
\hline
\hline
 Source &  Energy band &  $N_{\rm H_{\rm Gal}}$ & $N_{\rm H}$ &  $\Gamma$ & kT&  $\chi^{2}/\nu$ \\
     & (keV)  & ($10^{22}$ cm$^{-2}$)  &($10^{22}$ cm$^{-2}$) &  &   (keV)  &    \\
\hline
\hline
 HESS J1640--465 (\# 1) & 0.2--6.7 & 2.22  & 9.6$^{a}$ & 2.59$^{+1.03}_{-1.08}$& --  & 6.9/11 \\
\hline
 HESS J1640--465 (\# 2) & 0.2--5 & 2.18 & 3.41$^{+8.44}_{-3.41}$ & 1.84$^{+1.35}_{-1.10}$& -- &
4.3/8 \\  
\hline
 HESS J1640--465 (\# 3)& 0.2--3 & 2.19 & 0.42$^{+0.34}_{-0.22}$ & --  & 0.43$^{+0.37}_{-0.19}$$^{b}$  &
8.8/13 \\ 
\hline
HESS J1804--216 (\# 1) & 0.2--2 &  1.53 & -- & -- & 0.29$^{+0.18}_{-0.10}$$^{c}$ & 6.4/6 \\
\hline
 HESS J1813--178 (\# 1) & 0.2--7.5 &  1.89 & $9.67^{+9.62}_{-4.55}$ & $1.80^{+1.66}_{-0.97}$& -- 
& 32.8/38 \\
\hline
 HESS J1837--069 (\# 1) & 0.2--8.5 & 1.86  &$5.54^{+2.83}_{-1.81}$ & 1.86$^{+0.69}_{-0.46}$ & --
& 15.1/22\\
\hline
\enddata

\tablecomments{${^a}$ The column density is fixed to the \emph{ASCA} value (see Sugizaki et al. 
(2001));\\
${^a}$ kT is estimated modeling the data with a thermal emission from thin plasma (Raymond \& Smith 
1977);\\
${^c}$ kT is estimated modeling the data with a thermal bremsstrahlung emission.}
\end{deluxetable}

\clearpage

\begin{deluxetable}{lccccc}
\tablecolumns{6}
\tablecaption{Possible identifications of the sources detected within the XRT field of view of HESS 
J1837--069.}
\startdata
\hline
\hline
Source  &Catalog & \multicolumn{2}{c}{Coordinates}& $R$   & $K$ \\
    &  &  R.A.(J2000) &  Dec(J2000) &     &     \\
\hline
\# 1 & 2MASS/USNO-B1.0 & 18$^{h}$ 38$^{m}$ 03$^{s}$.26 & --06$^{\circ}$ 55$^{\prime}$ 39$^{\prime
\prime}$.5 & 17.6 & 10.7 \\
\# 2 & 2MASS/USNO-B1.0 & 18$^{h}$ 37$^{m}$ 50$^{s}$.00 & --07$^{\circ}$ 07$^{\prime}$ 31$^{\prime
\prime}$.4 & 13.9 & 9.20 \\
\# 3 & 2MASS/USNO-B1.0 & 18$^{h}$ 37$^{m}$ 41$^{s}$.34 & --07$^{\circ}$ 05$^{\prime}$ 14$^{\prime
\prime}$.4 & 17.0 & 13.2 \\
\# 5 & USNO-B1.0 & 18$^{h}$ 37$^{m}$ 02$^{s}$.68 & --06$^{\circ}$ 49$^{\prime}$ 25$^{\prime
\prime}$.2 & 18.0 & -- \\
\# 7 & 2MASS & 18$^{h}$ 37$^{m}$ 23$^{s}$.18 & --06$^{\circ}$ 52$^{\prime}$ 11$^{\prime
\prime}$.0 & -- & 13.2 \\
\# 8 & USNO-B1.0 & 18$^{h}$ 37$^{m}$ 58$^{s}$.93 & --06$^{\circ}$ 49$^{\prime}$ 24$^{\prime
\prime}$.7 & 18.5 & -- \\
\hline
\enddata
\end{deluxetable}

\clearpage

\begin{figure}
\plotone{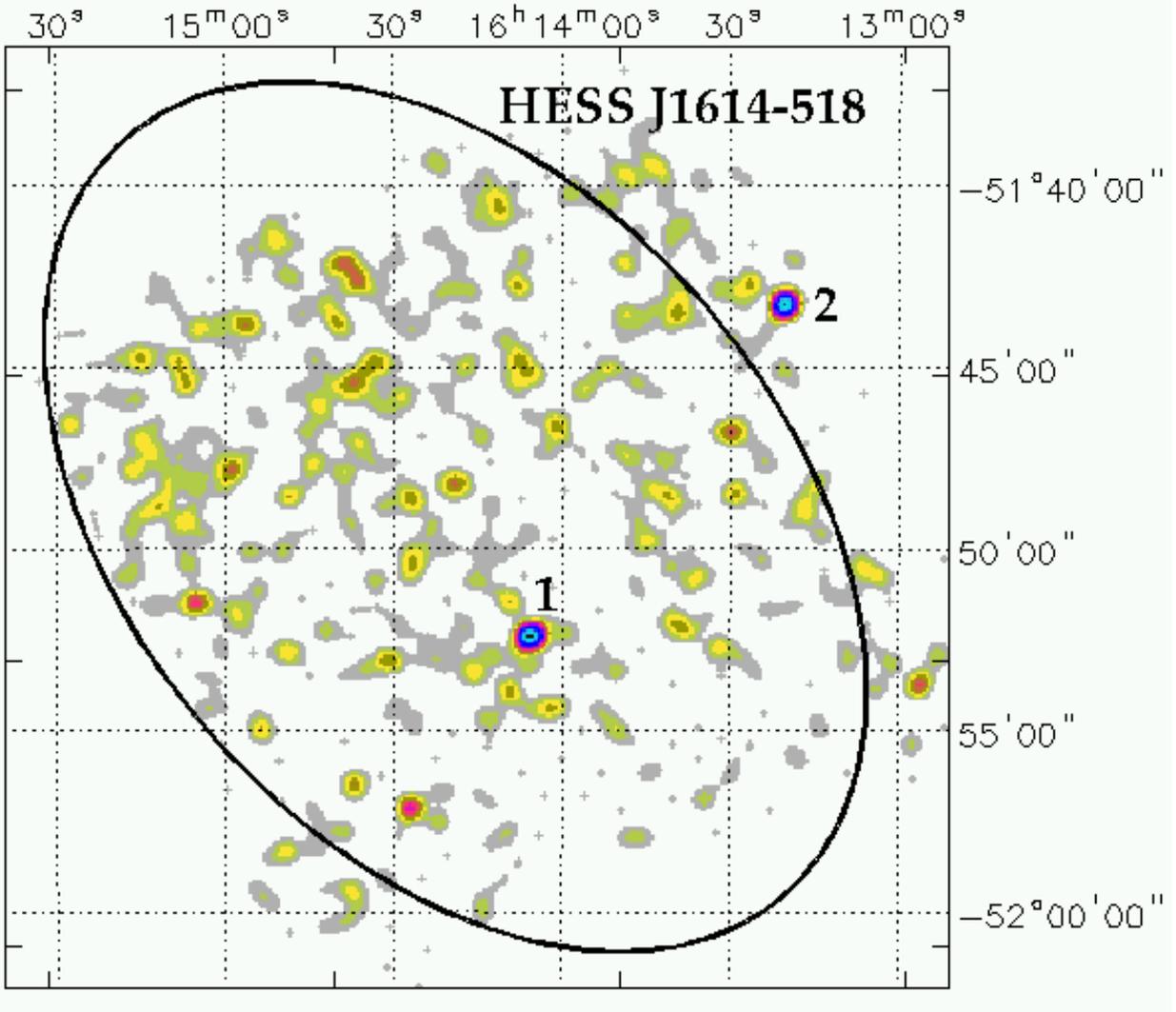}
\caption{XRT 0.3--10 keV image of the region surrounding HESS J1614--518. The ellipse 
represents the extension of the TeV source, while X-ray sources are labelled as in Table 2.}
\end{figure}

\clearpage

\begin{figure}
\plotone{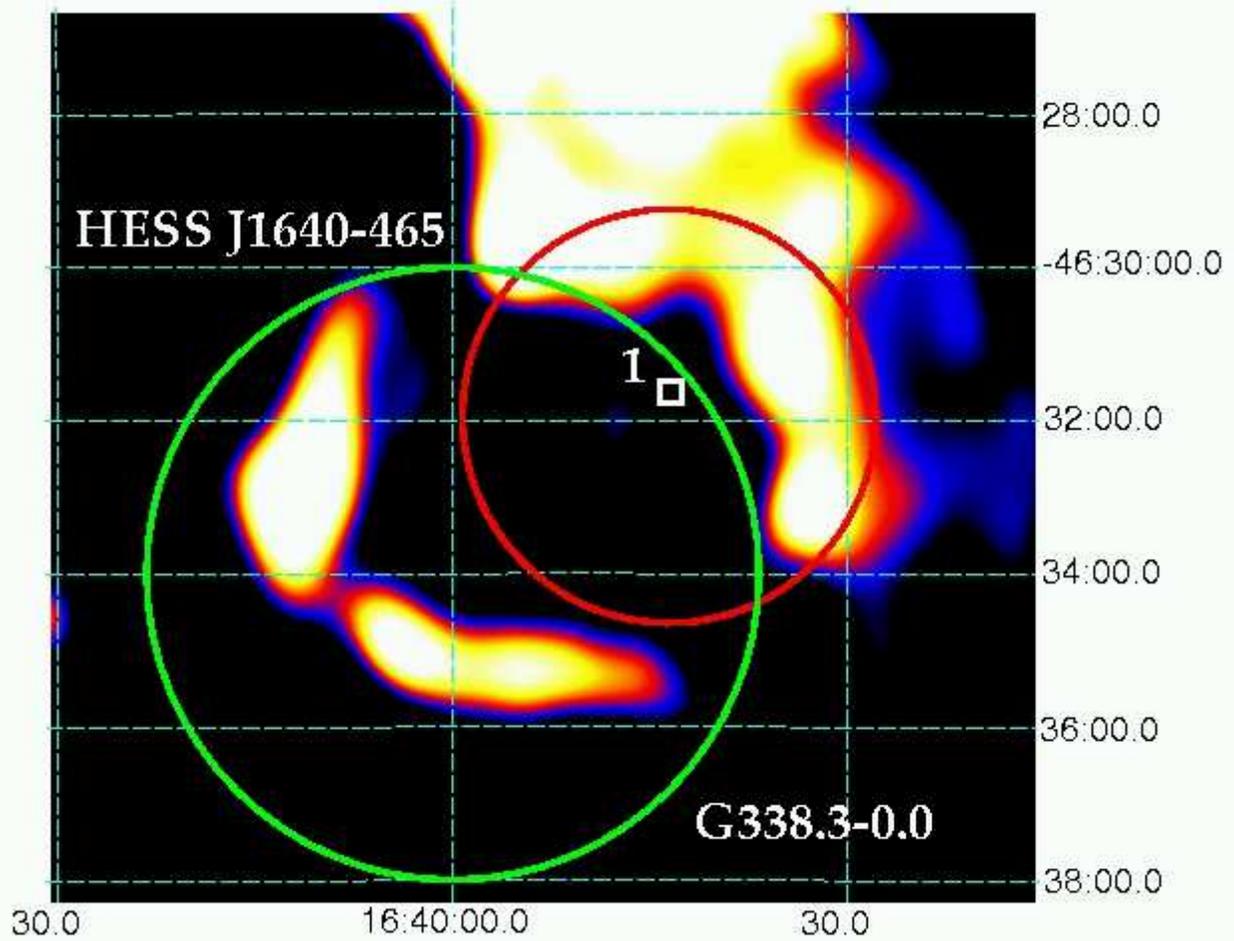}
\caption{MOST 843 MHz radio image of the region surrounding HESS J1640--465. The larger
circle (green) describes the 
position and extension of SNR G338.3--0.0 as given in Green (2004). The smaller circle (red) represents 
instead the 
extension of the TeV source. The position of the XRT Source 1 is given by a box.} 
\end{figure}

\clearpage

\begin{figure}
\plotone{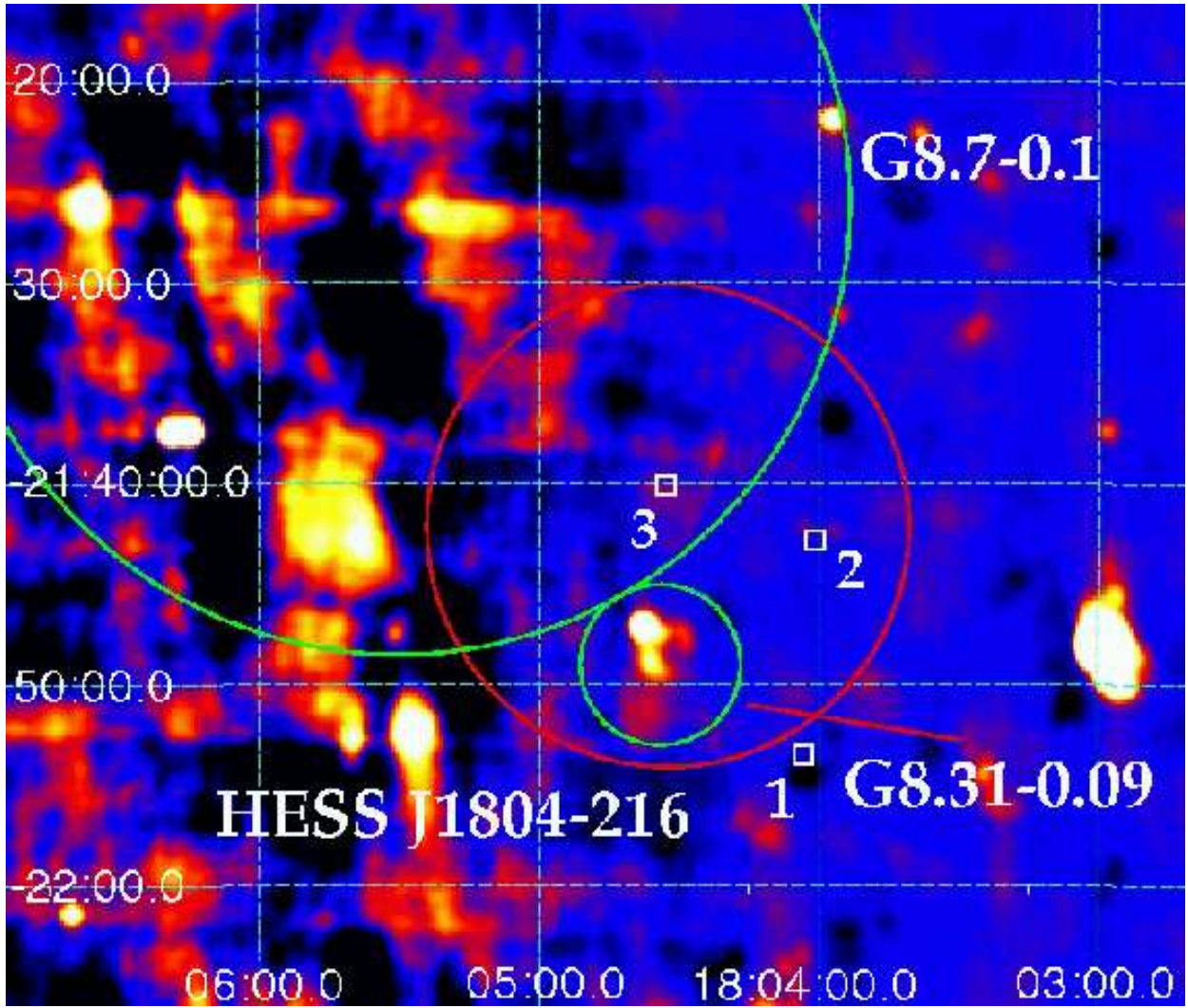}
\caption{NVSS 20 cm radio image of the region surrounding HESS J1804--216. The part of the circle 
(green) at the 
top encompasses the portion of SNR G8.7--0.1, which is inside the HESS extension. The smaller complete 
circle (green) indicates the position and extension of the new SNR detected by Brogan et al. (2006). The 
larger complete circle (red) instead represents the extensions of the TeV source. Here too, the positions 
of XRT sources are given by boxes.} 
\end{figure}

\clearpage

\begin{figure}
\plotone{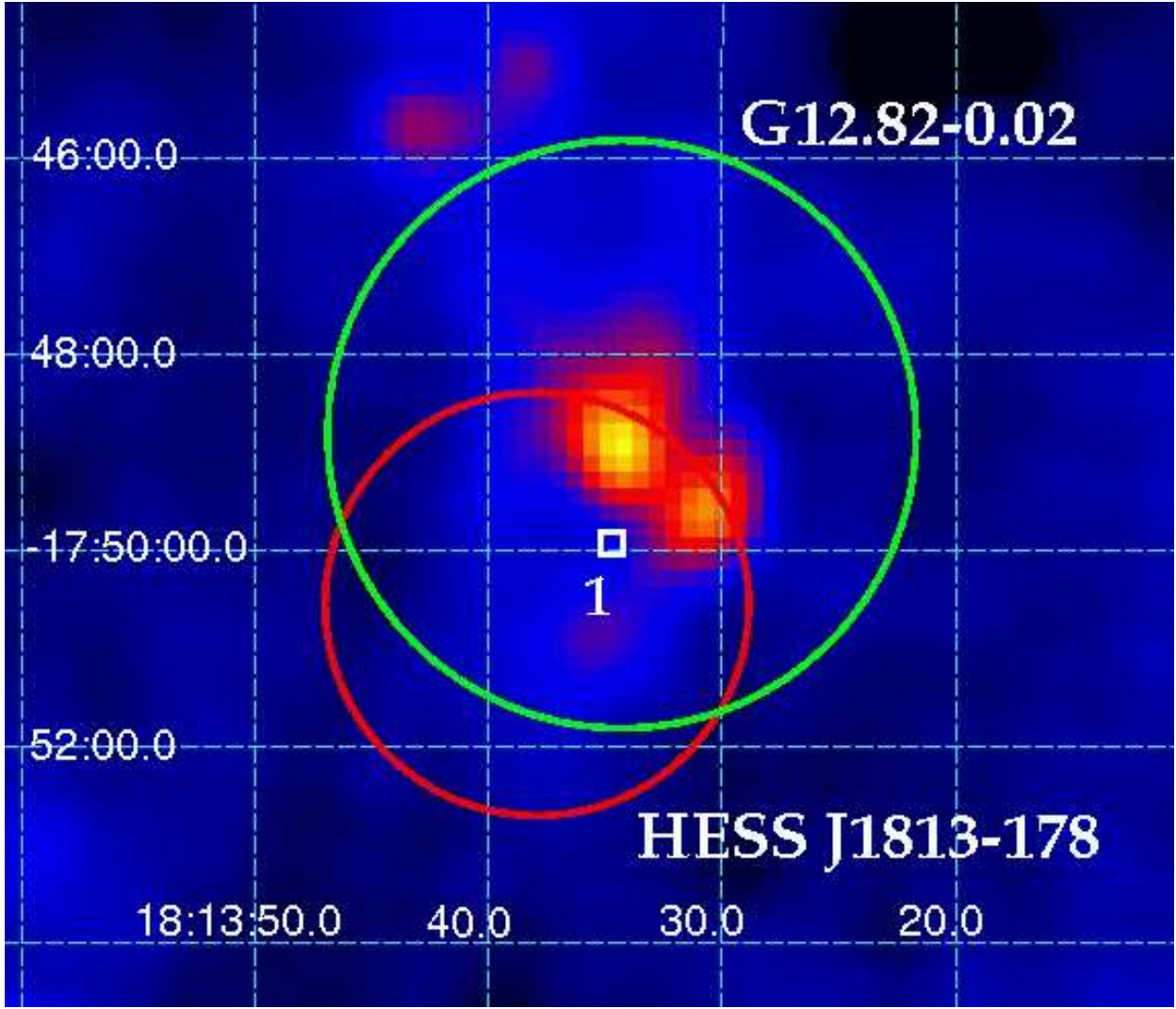}
\caption{NVSS 20 cm radio image of the region surrounding HESS J1813--178. The larger (green) circle 
describes the 
position and extension of SNR G12.82--0.02 as given in Brogan et al. (2005). The smaller circle 
(red) represents 
instead the extension of the TeV source. The position of the XRT Source 1 is given by a box.} 
\end{figure}

\clearpage

\begin{figure}
\plotone{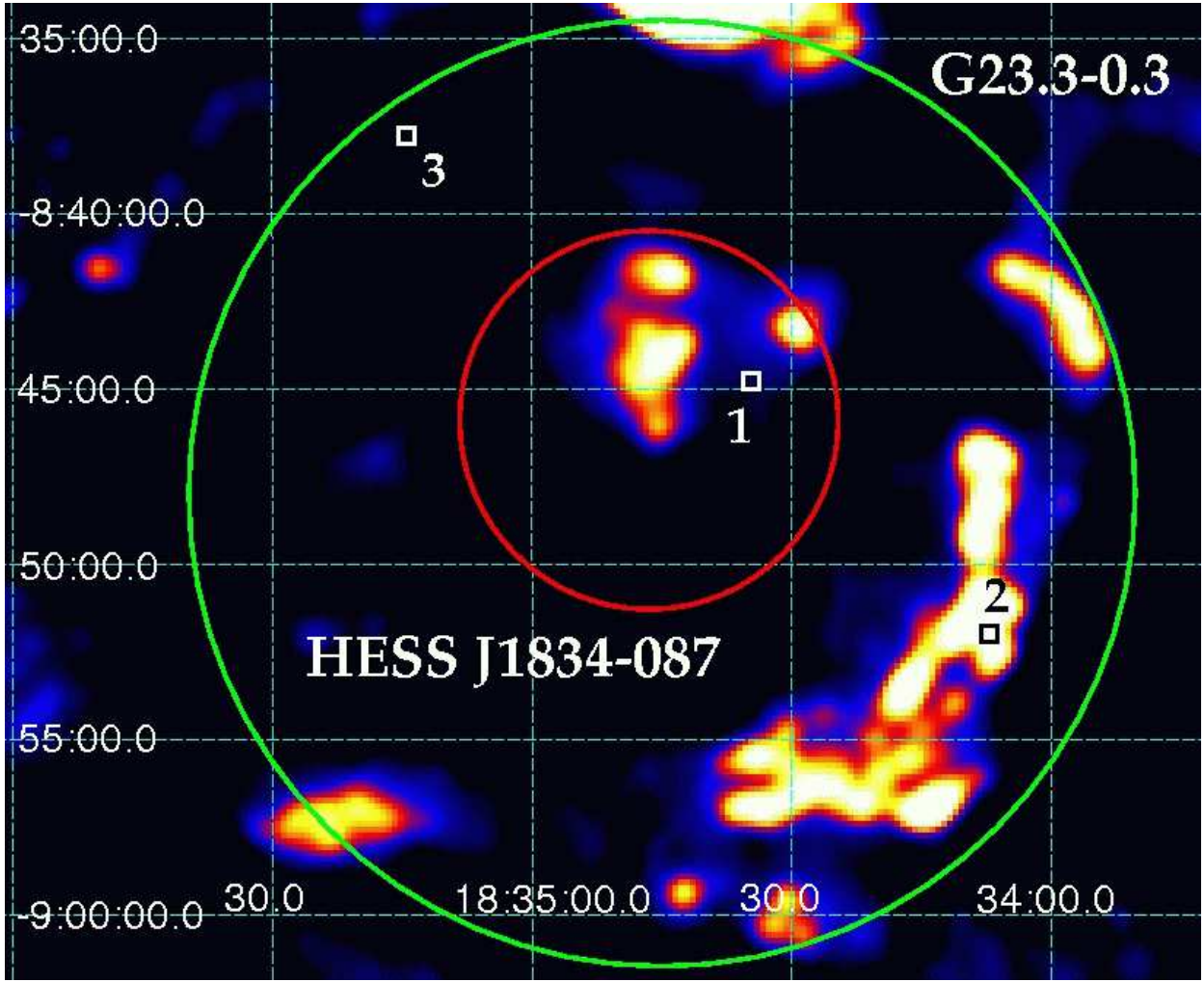}
\caption{NVSS 20 cm radio image of the region surrounding HESS J1834--087. The larger circle 
(green) describes the
position and extension of SNR G23.3--0.3 (W41) as given in Green (2004). The smaller circle 
(red) represents
instead the extension of the TeV source. The positions of the XRT sources are given by boxes 
(see Table 2).}
\end{figure}

\clearpage

\begin{figure}
\plotone{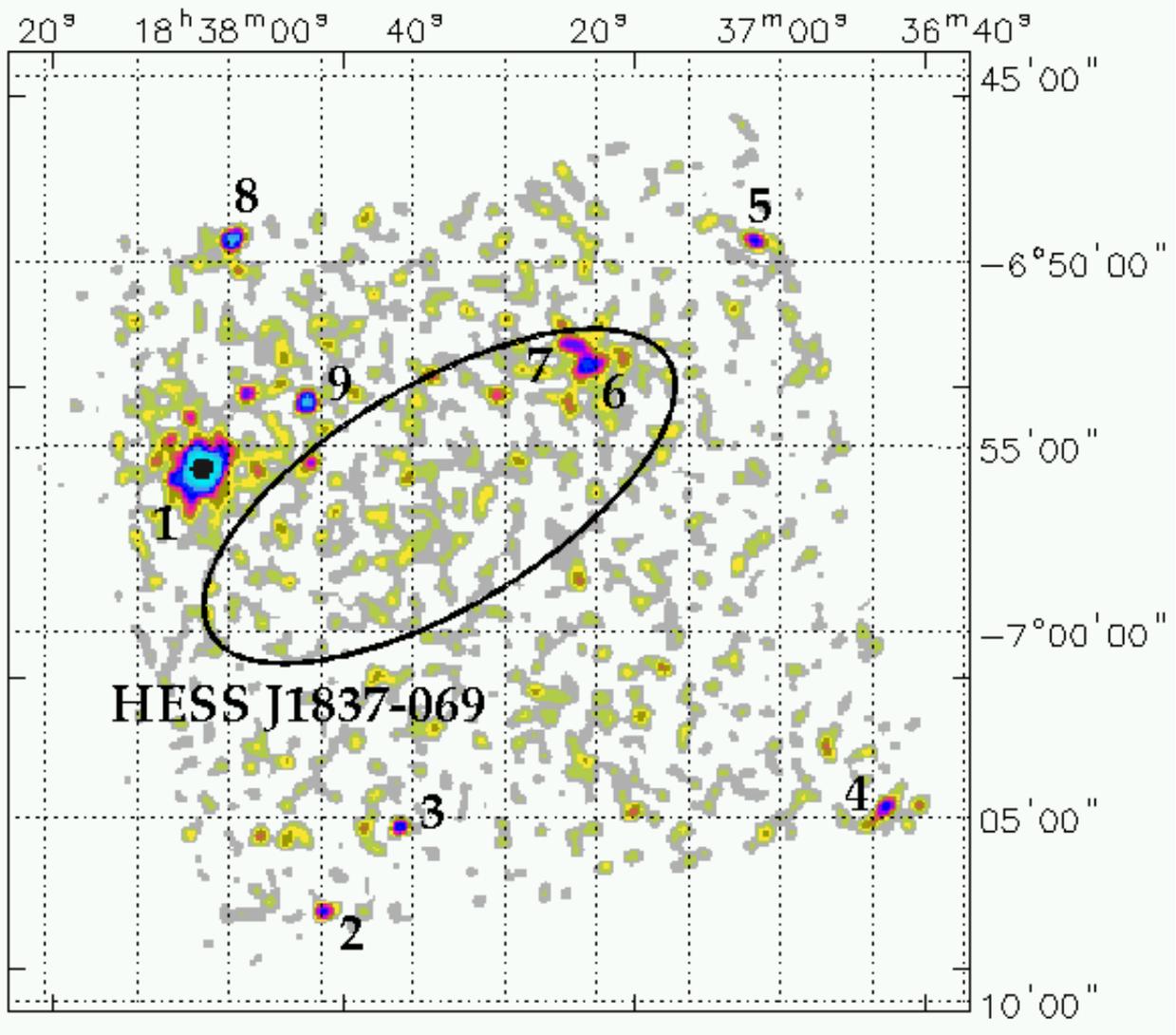}
\caption{XRT 0.3--10 keV image of the region surrounding HESS J1837--069. The ellipse represents the 
extension of the TeV source, while detected objects are labelled as in Table 2.}
\end{figure}

\end{document}